\documentclass[a4paper]{jpconf}
\usepackage{amsmath,slashed,amssymb,epsfig,amsthm,bm,graphicx}
\usepackage{caption}

\usepackage{color}
\definecolor{darkgreen}{rgb}{0,0.35,0}

\usepackage{tikz}
\usetikzlibrary{matrix}
\usepackage{enumerate}

\def\HH{\mathsf{H}} 
\def\HHG{\mathsf{H_G}} 
\def\HHF{\mathsf{H_F}} 
\def\PP{\mathsf{P}} 
\def\PPF{\mathsf{P}_{\hspace{-0.03cm}\mathrm{F}}} 
\def\PPG{\mathsf{P}_{\!\mathrm{G}}} 

\def\BH{\mathrm{BH}} 
\def\Tr{\mathrm{Tr}}

\def\eps{\varepsilon} 
\newcommand{\ket}[1]{| #1 \rangle} 
\newcommand{\bra}[1]{\langle #1 |} 

\newcommand{\avg}[1]{\langle #1 \rangle}

\begin{document}
\title{Two arguments for more fundamental building blocks}

\author{Alfredo Iorio}

\address{Faculty of Mathematics and Physics, Charles University, V Hole\v{s}ovi\v{c}k\'ach 2, 18000 Prague 8, Czech Republic}

\ead{alfredo.iorio@mff.cuni.cz}

\begin{abstract}
We present two lines of reasoning, leading to elementary constituents more fundamental than the ones we know. One such arguments is new, and based on the holographic maximal bound for the number of degrees of freedom of any system. In this case, both matter and space are emergent. The other argument is old, and was given by Richard Feynman as a possible explanation of why analog systems do describe the same physics. The former argument naturally points to a solution of the information paradox. The latter argument elevates analogs from mere curiosities, to reliable tests of fundamental theories. Amusingly, the names given to this fundamental level, both by Feynman and by some of the modern quantum gravity researchers, e.g., Jacob Bekenstein, resemble each others: ``$X$ons'' (Feynman) vs ``level $X$'' (Bekenstein).
\end{abstract}

\section{Introduction}

Here we elaborate on the work of \cite{ACQUAVIVA2017317}, where it is proposed that, at our energies, the fields and spacetimes they live in are entangled. This entanglement is there because both, matter and space, emerge from the dynamics of the same more fundamental objects, whose existence can be inferred from the celebrated upper bound on the entropy of any system, conjectured by Bekenstein  \cite{Bekenstein1981,Bekenstein2014}. Quoting Feynman \cite{feynman2006feynman}, and paraphrasing Bekenstein \cite{Bekenstein2003}, here we call those objects ``$X$ons''.

The most noticeable result of \cite{ACQUAVIVA2017317} is that the evaporation of a black hole ($\BH$) can only lead to an information loss, in the sense that, in general, there is a nonzero entanglement entropy associated to the final products of the evaporation. In what follows, we recollect that result, next to suggestions on how to realize the model on certain analog systems. This gives us the chance to discuss briefly analogs in general, and to offer a reflection on Feynman's famous (but not well-known...) lecture ``Electrostatic Analogs'', that is chapter twelve of the second volume of \cite{feynman2006feynman}.

The structure of the paper is as follows. In Section \ref{BekensteinXons} we explain why the Bekenstein bound could mean that $X$ons exist, we present the quasiparticle picture stemming from this, and we show that it could be the key to the solution of the information-loss puzzle. In Section \ref{FeynmanXons} we discuss how the above scenario could be realized in analogs, then we face the issue of analogs in general, and we close that Section by reflecting on Feynman's conclusions on why analogs do work, leading him to evoke $X$ons, as the ultimate elementary constituents of matter\footnote{As we shall show later, Feynman went very close to postulate $X$ons as elementary constituents of \textit{space}, as well as matter. Indeed, it was the analysis of intimate properties of space itself, that lead him to conjecture the existence of these objects. Nonetheless, there is no clear statement he makes in this respect.}. That striking argument is simply based on the observation of what we daily have around. It was given decades before the much more sophisticated arguments of quantum gravity. The final Section is devoted to our conclusions.

\section{The new story: Bekenstein's $X$ons and the unavoidable information loss}\label{BekensteinXons}

Bekenstein proposed \cite{Bekenstein1981} that the entropy of any system, in a given volume $V$, can never be bigger than the entropy of the
$\BH$ whose event horizon coincides with the boundary of $V$
\begin{equation}\label{eq:BekensteinBound}
	S \leq S_{BH} \,.
\end{equation}
In the proper units (and forgetting $\log$ corrections), one has \cite{Bekenstein-1973} $S_{BH} = A/4$, where $A = \partial V$. The implications of the \textit{holographic} aspects of (\ref{eq:BekensteinBound}) go far and wide, see, e.g., \cite{tHooft1993,Susskind1995,Bousso1999,Bousso2002}. We focus here, instead, on the fact that (\ref{eq:BekensteinBound}) is a universal \textit{upper bound}.

That formula means that the Hilbert space associated to anything contained in $V$ has finite dimension $\sim e^{S_{BH}}$. The world at our energy scales is well described by fields and by the space they live in. Fields, as we know them, have infinite-dimensional Hilbert spaces, to which one should add the degrees of freedom surely carried by (the quanta of) space itself. How can then be that the \textit{ultimate} Hilbert space, that must include \textit{all} degrees of freedom, is not only separable, like for a single harmonic oscillator, but it is actually finite dimensional?

This simple logic points to the existence of something new, more fundamental. Something making both, fields and space. Objects in terms of which particles like the electron, the muon, etc. are, in fact, \textit{quasi-particles}, emerging from the interaction with some sort of \textit{lattice}, whose emergent picture is space itself. Inspired by Feynman (see \cite{feynman2006feynman} and later here) we call these objects: $X$\textit{ons}. To access the $X$ons one would need resolutions of the order of the Planck length, which might be not only technically unfeasible, but actually impossible (see, e.g., \cite{Doplicher1995}).

\subsection{The universal quasiparticle picture}

Emergent, nonequivalent descriptions of the same underlying dynamics are ubiquitous in quantum field theory (QFT) \cite{Haag1992, Dirac1966, Umezawa1982}, as, in general, the vacuum has a nontrivial structure \cite{Milloni2013} with nonequivalent ``phases''. That is, for a given basic dynamics one should expect several different Hilbert spaces, representing different ``phases'' of the system with distinct physical properties. Distinct excitations play the role of the elementary excitations for the given ``phase'' \cite{Umezawa1993}, but their general character is that of the quasiparticles of condensed matter \cite{Landau8,Landau9}. Noticeably, in theories with certain \textit{dualities}  \cite{Montonen1977,Castellani2016}, elementary and collective excitations are interchangeable pictures. On this we also comment in Section 3.

What we add here to that QFT picture are, essentially, two important ingredients:
\begin{itemize}
  \item the finiteness of the degrees of freedom, forcing to interpret fields as emergent;
  \item the necessity to include the spacetime into such emergent  picture\footnote{That gravity could be an emergent phenomenon is an old idea of Sakharov \cite{Sakharov1967,Visser2002}, with many present ramifications, such as, e.g., the ``world crystal'' model of Kleinert \cite{Kleinert1987}, the hydrodynamics approach of Jacobson \cite{Jacobson1995}, and many more, see, e.g., \cite{Baez1999,Lombard2016,Oriti2014}. Proposals of particular interest for our discussion are that of \cite{VanRaamsdonk2010}, where spacetime emerges from the quantum entanglement between the actual fundamental degrees of freedom, see also \cite{Cao2016}, while in ``quantum graphity'' \cite{Konopka2008}, fundamental degrees of freedom and their interactions are represented by a complete graph with dynamical structure.}.
\end{itemize}
Taking this view, the continuum of fields and space is then only the result of a limiting process. In general, there must be (many!) microscopic configurations of the $X$ons giving raise to the \textit{same} emergent geometry, but to \textit{different/non-equivalent} field configurations.

Let us now summarize a possible way to formalize the above, for more details see \cite{ACQUAVIVA2017317}. If $\HH$ indicates the fundamental Hilbert space of the $X$ons, what discussed above means that there is some mapping, $\PPG: \HH \mapsto \HHG$, which assigns to a microscopic state in $\HH$ the corresponding ``classical geometry''. On the other hand, there must be some mapping, $\PPF:\HH \mapsto \HHF$, which extracts the ``field content'' of a state in $\HH$.  The states of the $\HH$ can be interpreted as states with some classical geometry, $g^{(a)}$, via the mapping $\PPG$, and states of the quantum field, $\phi$, via the mapping $\PPF$:
\begin{center}
\begin{tikzpicture}
  \draw (0,0) node {$\ket{\psi}\in\HH$};
  \draw[->] (-0.5,-0.3)--+(-1.5,-0.5) node[pos=0.75,above=3pt] {\scriptsize $\PPG$};
  \draw[->] (0.3,-0.3)--+(1.5,-0.5) node[pos=0.75,above=3pt] {\scriptsize $\PPF$};
  \draw (-2, -1.1) node { $\ket{g^{(a)}}\in\HHG$};
  \draw (2, -1.1) node { $\ket{\phi} \in \HHF$};
\end{tikzpicture}
\end{center}
In other words, the Hilbert space $\HH_G$ represents the states which appear as classical spacetimes, on the emergent level, while $\HH_F$ is the Hilbert space representing states that appear as states of quantum fields, on the emergent level. Henceforth, the generic state in $\HH$ can be labelled by the values of the coarse-grained variables, $\ket{\psi}=\ket{g^{(a)}, \phi}$.

We can also split $\HH$ into a direct sum of the subspaces $T_{(i)}$,
\begin{equation}\label{Hfirst}
  \HH = \bigoplus_{i=1}^{N_T} T_{(i)}, \qquad \dim \HH = N_T \, N,
\end{equation}

where each $T_{(i)}$ has a fixed dimension $N$ and a specific ``topology'', i.e., a specific distribution of fundamental degrees of freedom among the $\ket{g^{(a)}}$ and the $\ket{\phi}$, and a specific arrangement. $N_T$ is then the number of different topologies. By assumption, each $T_{(i)}$ has a structure
\begin{equation}\label{T(i)}
  T_{(i)} = \HH_{\mathrm{G}}^{p_i} \otimes \HH_{\mathrm{F}}^{q_i}, \qquad p_i\,q_i = N,
\end{equation}
where $\HH_{\mathrm{G}}^{p}$ ($\HH_{\mathrm{F}}^q$) is a Hilbert space of dimension $p$ ($q$) representing possible microscopic states of the geometry (fields). With this in mind, the generic state $\ket{\psi}\in\HH$ can be written as
\begin{equation}
  \ket{\psi} = \bigoplus_{i=1}^{N_T} \sum_{I=1}^{p_i} \sum_{n=0}^{q_i-1} c^{(i)}_{In} \ket{I_i} \otimes \ket{n_i},
\end{equation}
where the vectors $\ket{I_i}$ and $\ket{n_i}$ form a basis of $\HH_{\mathrm{G}}^{p_i}$ and $\HH_{\mathrm{F}}^{q_i}$, respectively, and $c^{(i)}_{In}$ are numerical coefficients.

By denoting with ${\PP}_{(i)} : \HH \mapsto T_{(i)}$ a projector onto $T_{(i)}$, the squared norm of ${\PP}_{(i)}\ket{\psi}$
is the probability $p_{(i)}$ of finding the system in the state with the topology $T_{(i)}$,  $p_{(i)} = \| {\PP}_{(i)}\ket{\psi}\|^2$.

Importantly, the generic state of $T_{(i)}$ has entanglement between the geometry and the field, in the sense that its decomposition reads
\begin{equation}
  {\PP}_{(i)} \ket{\psi} = \sum_{I,n} c^{(i)}_{In} \ket{I_i} \otimes \ket{n_i} \,.
\end{equation}
The associated density matrix, representing the state of the field, is
\begin{equation}\label{eq:rho-i}
  {\rho}_{(i)} = \Tr_{\HH_{\mathrm{G}}^{p_i}} \ket{\psi}_{i} \bra{\psi}_i ,
\end{equation}
where we first define the normalized state $\ket{\psi}_i = p_{(i)}^{-1/2} {\PP}_{(i)}\ket{\psi}$, and then trace over the degrees of freedom of the gravitational field. Correspondingly, the entropy of geometry-fields entanglement, for a given topology of the lattice is then the usual expression
\begin{equation}\label{eq:S-i}
  S_{(i)} = - \Tr_{\HH_{\mathrm{F}}^{q_i}} {\rho}_{(i)} \ln {\rho}_{(i)} \,.
\end{equation}
As it is impossible to distinguish the space corresponding to different topologies of the lattice, the expected value of the entanglement between the fields and the geometrical degrees of freedom is
\begin{equation}\label{eq:Savg}
\langle S \rangle = \sum_i p_{(i)} S_{(i)} \,,
\end{equation}

This picture needs to be compared to the standard QFT picture, recalled earlier, of the non-equivalent field configurations, or ``phases''. It is very tempting to consider, in particular, the Thermo-Field Dynamics (TFD) \cite{Takahashi1996} framework, with its tilde degrees of freedom playing the role of (a coarse-grained and effective description of) the degrees of freedom of the geometry to trace away. Indeed, the vacuum of TFD can be written as \cite{Takahashi1996}
\begin{equation}\label{eq:TFDvacuum}
	\ket{0 (\theta)} = \sum_n \sqrt{w_n} \ket{n, \tilde{n}} ,
\end{equation}
where $\theta$ is a physical order parameter, such as temperature, magnetization, etc., (but also acceleration, surface gravity, etc., in gravitational contests  \cite{IORIO2001234,Iorio2004}) labeling the different ```phases'', $w_n$ are probabilities such that $\sum_n w_n =1$, and the states $\ket{n, \tilde{n}}$ (infinite in number) are the components of the condensate, each made of pairs of $n$ quanta and their $n$ mirror counterparts ($\tilde{n}$). Therefore, such vacuum is clearly an entangled state. Notice that \cite{Umezawa1982}
\begin{equation}\label{eq:TFDinequiv}
	\bra{0(\theta)} 0 (\theta') \rangle \to 0 ,
\end{equation}
in the field limit, that formalizes the inequivalence we have discussed. Notice also that, if one fixes $\theta$, there is no unitary evolution to disentangle the vacuum, as the interaction with the environment and non-unitarity are the basis for the generation and the stability of such entanglement \cite{Iorio2004}.

The expected value of \textit{field's} observables, $O$, are obtained by tracing away the \textit{mirror} modes, $\tilde{n}$. In the TFD formalism this corresponds to taking the vacuum expectation value over the vacuum (\ref{eq:TFDvacuum})
\begin{equation}\label{eq:TFDOavg}
\langle O \rangle \equiv \langle 0(\theta) | O | 0 (\theta) \rangle = Z^{-1} (\theta) \sum_n e^{\theta E_n} \langle n | O | n \rangle  \,,
\end{equation}
where $Z$ is a partition function.

In particular, there is always an \textit{entanglement entropy} associated to any field, given by, e.g.,
\begin{equation}\label{eq:TFDentropy}
\langle S \rangle = \sum_k \left[ n_k \ln n_k + (1 - n_k) \ln (1 - n_k) \right] \,.
\end{equation}
where $n_k = \langle N_k \rangle$ is the expected value of the number operator for the given (fermionic, in this example) mode $k$. The analogy of (\ref{eq:TFDentropy}) with (\ref{eq:Savg}) and (\ref{eq:S-i}) is stronger, if we think that in TFD the process of taking statistical averages through tracing is replaced, by construction \cite{Takahashi1996}, by taking vevs over the vacuum (\ref{eq:TFDvacuum}).

In this comparison, the mirror (tilde) image of the field mimics the effects of the entanglement with space where the field lives, even when the space is flat. This happens on a level that is both emergent and effective. This would have far reaching consequences, surely worth a serious exploration. For instance, the entanglement entropy associated to any field, would never be zero. Furthermore, this would explain why quantizing gravity as we quantize the matter fields, cannot make much sense.

\subsection{Effects on black hole evaporation}

When applied to $\BH$ evaporation, the immediate consequence of the above, is that it is impossible that after the evaporation we can retrieve the very same ``phase''  we had before the $\BH$ was formed. Hence, the information associated to the quantum fields before the formation of the $\BH$ is, in general, lost after the $\BH$ has evaporated, due to the fields-geometry entanglement.

Even when the emergent spaces, before the formation and after the evaporation, are the same, say Minkowski, the emergent fields belong, in general, to non-equivalent Hilbert spaces. Therefore, even assuming unitary evolution at the $X$ level, the initial and final Hilbert spaces of fields cannot be the same. There is always a \textit{relic} field-geometry entanglement entropy.

We shall now try to go all the way to actually computing these relic entropies, for an idealized scenario of $\BH$ evaporation, that we now describe.

We might think that initially matter, described by a quantum field, is in an almost flat space. Such matter then collapses\footnote{The mechanism for $\BH$ formation could be less immediate when the spacetime is of lower dimension. For instance, in three dimensions, although a solid and well known $\BH$ solution of Einstein gravity exists \cite{BTZ1992}, its degrees of freedom are of topological origin, i.e., due to specific boundary conditions (see, e.g.  \cite{Carlip:1998uc}), and collapse is not as immediate as in four dimensions.}, and eventually forms a $\BH$ of mass $M_0$. In our picture, the $\BH$ evaporates in a discrete manner, and we take each emitted quantum with the same energy $\eps$, so that $M_0 = N_G \,\eps$ for some integer $N_G$. At the end of the evaporation, the space becomes almost flat again, but the field is in a different state with $N_G$ quanta.

\begin{table}
\caption{\label{3cases} Three possible choices of the of the parameters, for two topologies, $N_T = 2$, thirty geometries, $N_G = 30$, and no ``field degeneracy'', $R_F = 1$. }
\begin{center}
\begin{tabular}{|c|c|c|c|c|c|c|}
  \hline
{\rm case} & $i$ & $R^i_G$  & $p_i=30 R^i_G$ & $q_i$ & $N=p_i q_i$ & $dim \HH = N_T N$ \\ \hline
I   & 1 & 1  & 30  & 200 & 6000 & 12000\\
    & 2 & 5  & 150 & 40  &  &  \\ \hline
II  & 1 & 2  & 60  & 200 & 12000 & 24000 \\
    & 2 & 10 & 300 & 40  &  & \\ \hline
III & 1 & 4  & 120 & 200 & 24000 & 48000 \\
    & 2 & 20 & 600 & 40  & &  \\
  \hline
\end{tabular}
\end{center}
\end{table}

Putting together (\ref{Hfirst}) and (\ref{T(i)}), the Hilbert space can be written as
\begin{equation}
\HH = \bigoplus_{i=1}^{N_T} \HH_G^{p_i} \otimes \HH_F^{q_i}   \label{eq:hilbert-space}
\end{equation}
where $N^T$ is, as said, the number of topologies, and we now introduce measures, $R_F$s,$R_G$s, of the ``degenaracies'', i.e.,
\begin{eqnarray}
  p_i &=& N_G\,R_G^i.
\end{eqnarray}
with $N_G$ classical geometries available (they represent the $\BH$ with mass $M^{(a)} = a\,\varepsilon$, where $a=0,1, \dots N_G-1$), and each classical geometry can be realized by $R_G^i$ microstates, while
\begin{eqnarray}
  q_i &=& N_F\,R_F^i.
\end{eqnarray}
that is, each emergent field state can be realized by $R_F^i$ indistinguishable microstates.

When the $\BH$ starts to evaporate, we assume continuous unitary evolution of the state
in $\HH$ but take the ``snapshots'' of the system, when the expected values are
\begin{equation}
  \avg{M} = (N_G-1-k)\,\eps, \quad \avg{n} = k,
\end{equation}
where $k$ acquires discrete values $k = 0, 1, \dots N_G-1$. Here $k=0$ corresponds to $\BH$ of maximal mass $M_0=M^{(N_G-1)}=(N_G-1)\eps$ and vacuum outside
the $\BH$. That does not necessarily mean that the Hilbert space for the field outside
the $\BH$ has dimension 1, because in our model such dimension depends on the topology. However, since there is only one vacuum
state in all topologies, the field is disentangled from the geometry. Hence, our starting point coincides with the starting point of the standard analysis of
Page \cite{Page1993a}: expected entanglement vanishes at the beginning of the evaporation.

At the $k$-th step, the $\BH$ already emitted $k$ quanta of the field, so its mass decreased by the value $k\,\eps$, while the field is in the state with $k$ quanta outside. When the $\BH$ is fully evaporated for $k=N_G-1$, the field is in the state with $N_G-1$ quanta.

\begin{figure}
  \centering
  \includegraphics[width=\textwidth]{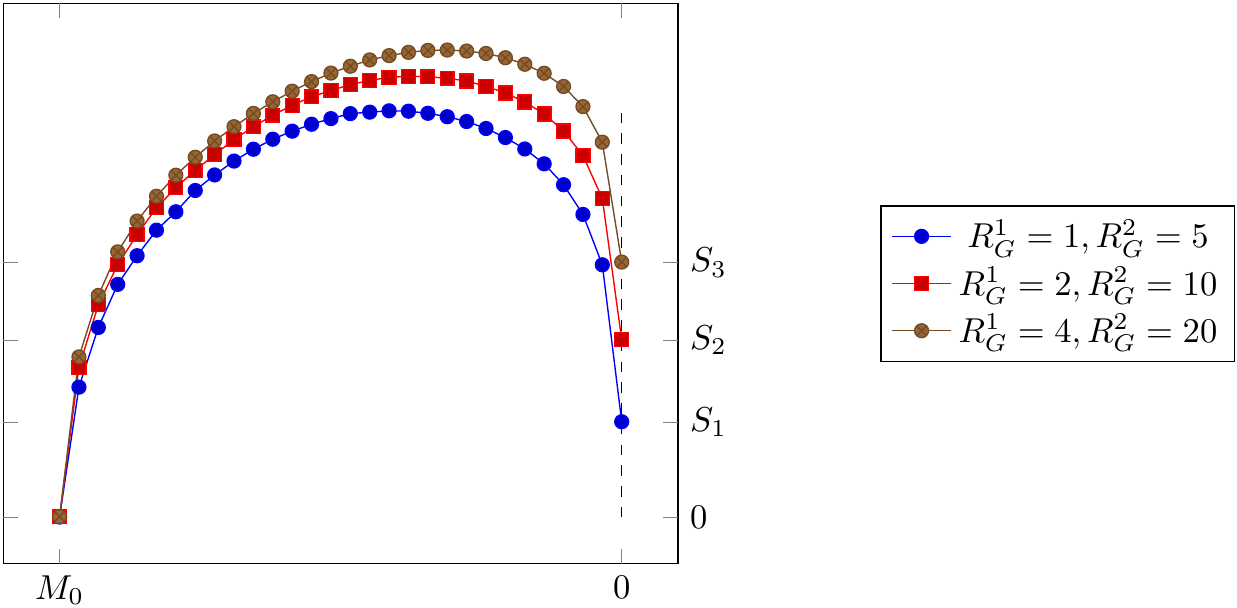}
  \caption{Field-geometry entanglement entropy, as a function of the decreasing mass of the evaporating $\BH$. The initial and final points of this curve are in exact correspondence with the initial and final points of the Page curve. The plot here is for two topologies and three cases. The more microscopic realizations of macroscopic classical geometries are allowed, the higher the residual entropies. Here: $S_1 = 0.77$, $S_2 = 1.43$, $S_3 = 2.06$. Taken from \cite{ACQUAVIVA2017317}.}
  \label{fig:page-modified_2}
\end{figure}

The actual analytic computations of the entanglement entropy are a heavy tall, so in \cite{ACQUAVIVA2017317} we proceed with easier (but not trivial...) numerical computations. The case we
want to report here is for the following choice of $N_G = 30$, $N_T=2$, and $R_F^i=1$, for each topology. We present in Table \ref{3cases} three cases, $I$, $II$, and $III$, and plot in Fig. \ref{fig:page-modified_2} the corresponding entanglement entropies, as functions of the discrete parameter $k$.

As can be seen from the figure, the residual entropies are never zero, and are given by
\begin{equation}\label{relicS}
  S_1 = 0.77, \quad S_2 = 1.43, \quad S_3 = 2.06 \,,
\end{equation}
for the cases $I$, $II$, and $III$, respectively. As it must be, the more microscopic realizations of the same macroscopic geometry (i.e. the bigger the degeneracy $R_G$), the higher the relic entanglement entropy.

\begin{figure}
  \centering
  \includegraphics[width=0.8\textwidth]{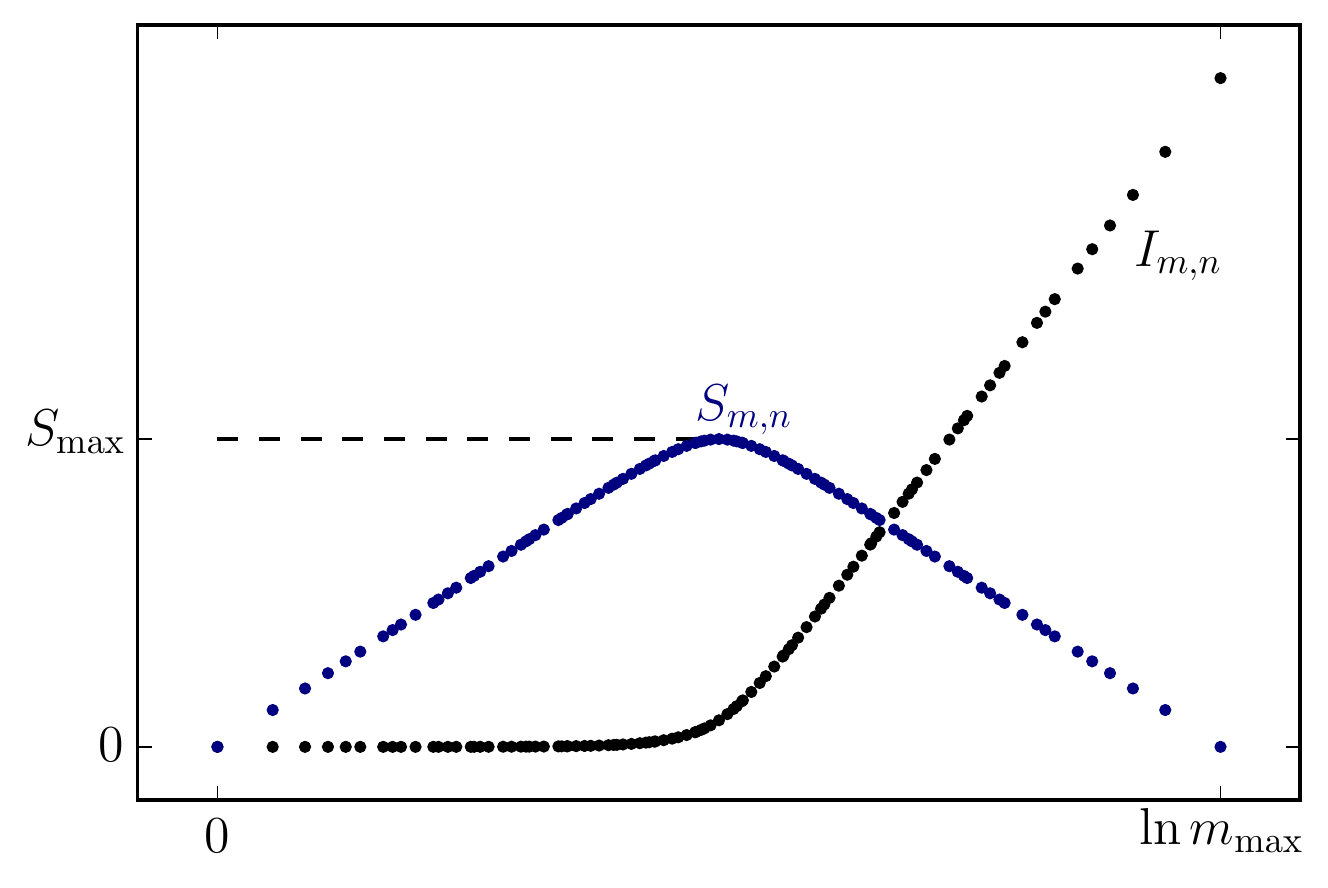}
  \caption{Field entropy of entanglement between modes inside the $\BH$ and modes of the radiation vs the dimension of the Hilbert space of the radiation, obtained in \cite{Page1993b} by Page via a unitary evolution. The point $m=0$ corresponds to the initial mass of the $\BH$, $M=M_0$, while $m_{max}$ corresponds a fully evaporated $\BH$, $M=0$.}
  \label{fig:page-original}
\end{figure}

The fact that, at the end of the evaporation, the entanglement entropy remains finite signals a dramatic departure from the {\it information conservation} scenario of the famous Page curve \cite{Page1993a,Page1993b}, presented here in Fig.  \ref{fig:page-original}. To fully appreciate this, we need to give the tools to compare the figures obtained within the quasiparticle non-unitaty picture of \cite{ACQUAVIVA2017317}, with that obtained in the fully unitary picture of \cite{Page1993a,Page1993b}.

In Page's scenario only fields are considered, and there is no explicit room for the degrees of freedom associated to the geometry. The Hilbert space represents the states of the field ``inside'' and ``outside'' the $\BH$ horizon. Thus, this is as for any standard bipartite system, $\HH = \HH_A^m \otimes \HH_B^n$,
where superscripts $m$ and $n$ indicate the dimension of the corresponding Hilbert space, so that $\dim \HH = m\,n$.
Then, picking an arbitrary fixed state $\ket{\psi_0} \in \HH$, and a random unitary matrix $U$, $U \ket{\psi_0}$ is a random state in $\HH$. By tracing away the subsystem $B$, to such state one associates the density matrix $\rho_A(U)$, and the corresponding entanglement entropy $S_{m,n}(U)$.
Averaging through $U$ we get the average entanglement entropy of the subsystem $A$, $S_{m,n} = \left\langle S_{m,n}(U) \right\rangle$ and the average \textit{information} contained in $A$, defined as
$I_{m,n} = \ln m - S_{m,n}$. It is found \cite{Page1993a}, see also \cite{Harlow2016}, that
\begin{equation}\label{eq:page information}
  I_{m,n} = \ln m + \frac{m-1}{2n} - \sum_{k=n+1}^{mn} \frac{1}{k} \,,
\end{equation}
for $m < n$.

These results, applied to $\BH$ evaporation, give the curves in Fig.\ref{fig:page-original}. There we see that, when the $\BH$ is formed, there is no Hawking radiation outside, hence, $n = 1$ and $m = \dim \HH$. $S_{m,n}$ is trivially zero. As the $\BH$ evaporates, $n$ increases while $m$ decreases, keeping $m \,n$ constant. Since the emitted photons are entangled with the particles under the horizon, $S_{m,n}$ increases, but only
up to, approximately, half of the evaporation process. There, the information stored below the horizon starts to leak from the $\BH$, so that $S_{m,n}$ decreases untill full evaporation,
hence $m = 1$ and $n = \dim \HH$ and $S_{m,n}$ returns to zero.

To conclude this Section, we may say that, even if one takes a conservative view, for which the $X$ons evolve unitarily, nonunitarity is unavoidable. Indeed,
\begin{itemize}
  \item the unitary evolution may as well be only formally possible, but physically impossible to spot, for a generalized uncertainty forbidding the necessary Planck scale localization/resolution (see, e.g., \cite{Doplicher1995});
  \item the emergent description of the evolution is that of the combined system gravity+matter, hence there is inevitably information loss, due to the relic entanglement of the matter field with the geometry;
  \item this description should apply also to standard nonunitary features of QFT, and we evoke here the possibility that the tilde degrees of freedom of TFD could be interpreted as ``how the emergent fields see the degrees of freedom of geometry with which they are entangled''.
\end{itemize}
Notice that our description does allow for an arbitrary number of different fields, hence naturally includes the possibility of yet unknown kinds of matter.

\section{The old story: Feynman's $X$ons and the elevation of analogs' status}\label{FeynmanXons}

It is extremely important to bring the long lasting open questions on the information loss, and on many other fundamental issues, under the judgement of experiments. In fact, we believe that there is probably no way to solve these issues solely on the basis of logical consistency of no matter which theory. This is also witnessed by the fact that, over the years, many have changed their views on the information loss and related issues, including Stephen Hawking himself\footnote{I cannot refrain to report here a funny anecdote. At the keynote lecture of a conference in Vienna, Roger Penrose, in a large and packed lecture-hall, wanted to defend his persistent view that information loss indeed takes place during $\BH$ evaporation. As he wanted Stephen Hawking on his side he said ``I agree with the old Hawking, and by \textit{old} I mean the young one''.}.

\begin{figure}
  \centering
  \includegraphics[width=0.4\textwidth]{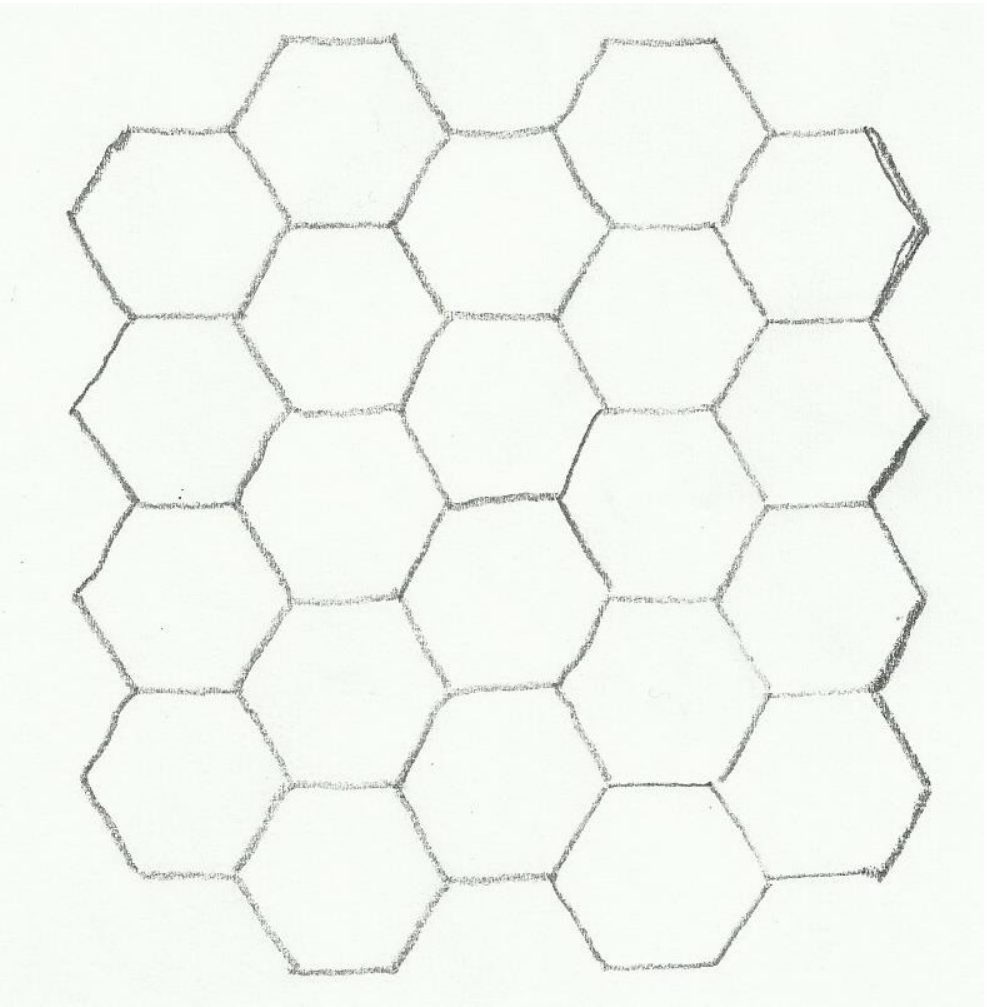}
\caption{An hexagonal lattice, common to many Dirac materials.}
  \label{Hexagons}
\end{figure}

For instance, it would be extremely interesting to perform an analog experiment on $\BH$ evaporation, to see which picture is correct, Page's or ours. We do not suggest here a specific set-up for the experiment, but suggest, instead, how certain scenarios, evoked in the quasiparticle picture of last Section, could find analog realizations in condensed matter. In fact, such analog scenarios inspired the theoretical model of \cite{ACQUAVIVA2017317} in the first place.

Examples of quasiparticles are many, some more, some less cogent as possible tests of our picture. Quasiparticles are the Cooper pairs of type II superconductors \cite{BSC1957,Altland2010} that are bosonic free excitations emerging from the basic dynamics of the electrons interacting with the lattice. Quasiparticles are the more recently discovered low energy excitations of graphene (see, e.g., the review \cite{CastroNeto2009}), that are massless Dirac pseudo-relativistic fermions (the matter $\psi$), emerging from the dynamics of electrons propagating in a carbon two-dimensional honeycomb lattice, whose emergent (continuum limit) description is that of a pseudo-Riemannian (2+1)-dimensional spacetime $g_{\mu \nu}$. Here the role of the $X$ons is played by carbon's electrons, ions and photons.

Indeed, we know from previous research \cite{hawkinggrapheneplb,hawkinggrapheneprd, IorioPais,IORIO2018265} (see also the review \cite{IorioReview}), that the emergence of intrinsic as well as extrinsic curvature in graphene can be used to probe the fundamental physics of the quantum Dirac field theory in the presence of a variety of spacetimes, or even quantum gravity scenarios \cite{IJMPD_Paper}.

\begin{figure}
  \centering
  \includegraphics[width=0.6\textwidth]{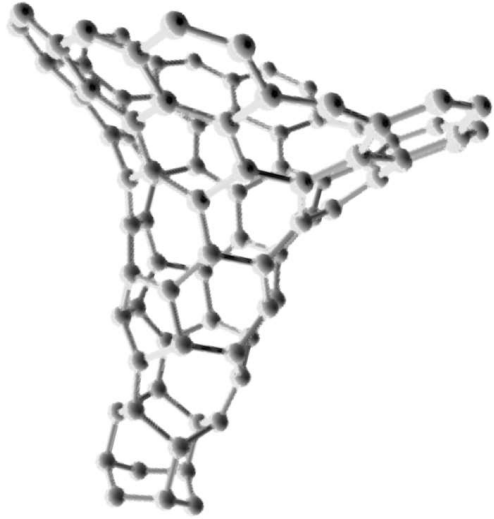}
\caption{A configuration with the necessary minimal number (six) of heptagonal defects, in order to have the intrinsic Gaussian curvature of a Beltrami pseudospehere \cite{Iorio_2013,Taioli_2016}. In the appropriate continuum limit, that is for a large number of atoms, and for large enough wavelengths of the $\pi$-electrons, this configuration realizes a spacetime conformal to a $\BH$.}
  \label{MiniBeltrami}
\end{figure}

\subsection{Black hole evaporation in graphene and graphene-like materials}

Let us start with an hexagonal lattice of ions, hold together by the $\sigma$ bonds. On this lattice $\pi$-electrons are the quasiparticles, whose properties are determined by a combination of the effects of the hexagonal geometry, and of the quantum interaction with the ions. One can associate  a ``flat world'', $(\eta, \psi)$, to the configuration of Fig. \ref{Hexagons}, that is free of inelastic defects. The world $(\eta, \psi)$ is  made of two components: the flat spacetime $\eta$, that is the emergent description of the lattice, and the Dirac field $\psi$, that is the emergent description of the $\pi$-electrons. In general, one should also add a $U(1)$ field that is the strain-induced gauge field with components, $A_x \sim u_{x x} + u_{y y}$ and $A_y \sim u_{x y}$, where $u_{ij}$ is the strain tensor, but for simplicity, we do not consider that here.

Then defects set in, according to specific arrangements \cite{Iorio_2013,Taioli_2016}, and make shapes like those of Fig.\ref{MiniBeltrami} , that, in certain specific limits, recreate conditions analog, through a conformal mapping, to those of a three-dimensional $\BH$ \cite{hawkinggrapheneplb,hawkinggrapheneprd}. In the example of Fig.\ref{MiniBeltrami} there are exactly 6 heptagonal defects, each carrying a Gaussian curvature $K_7 =  - \pi / 6$ \cite{Iorio_2013}.

In the limit of a large number of atoms (a necessary condition for the emergent description, along with wavelength of $\pi$-electron much larger than the lattice spacing), more and more defects, in the form of chains pentagons-heptagons, set in. Each such chain carries either a surplus of one heptagon, or even a net zero curvature. The role of these chains, usually referred to as \textit{scars}, is to distribute the curvature as evenly as possible on the surface. The emergent description of this is a ``$\BH$ world'', that we indicate with $(g^{(a)}_{\BH}, \psi_{\BH})$.

In a process that carries away the excess defects, that would mimic the $\BH$ evaporation through the running of the parameter $a$, one should arrive again at a ``flat world'', that is a membrane with overall zero curvature and zero torsion. Such flat world would still have $\eta$, as emergent description for spacetime, and some sort of flat space Dirac field as emergent description for the field.

What is interesting here is that, the final product of this process, in general, could have defects still trapped into its texture but, if arranged in specific manners, the emergent spacetimes are indistinguishable from $\eta$.

Three such examples are shown in Figs. \ref{585}, \ref{STW} and \ref{575757}. These defects configurations are routinely found in real materials \cite{ARAUJO201298}, and include the famous Stone–-Thrower-–Wales (STW) defect \cite{Thrower1969,STONE1986501}, shown in Fig. \ref{STW}.

They all have some nonzero local Gaussian curvature, at the location of the defect, but due to the arrangement of such defects, the curvature cannot propagate, giving an overall zero total curtavure. For Fig.\ref{585} one has $K_{tot} = 2 K_5 +  K_8 = 2 (\pi / 6)  - 2 \pi / 6 = 0$, for the STW defect of Fig.\ref{STW} one has $K_{tot} = 2 K_5 + 2 K_7 = 2 (\pi / 6) + 2 (- \pi / 6) = 0$, and, finally, for Fig.\ref{575757} one has $K_{tot} = 3 K_5 + 3 K_7 = 3 (\pi / 6) + 3 (- \pi / 6) = 0$. Furthermore, as is clear from the figures, these configurations do not introduce any nonzero Burger vector beyond the localized deformation, hence also torsion is absent from the emergent picture \cite{IORIO2018265}. Summarizing, the emergent spacetimes, in all three cases, are indeed $\eta$ spacetimes.

On the field side, though, the situation is each time different, becasue the spectrum of the $\pi$-electrons is actually modified by the presence of such defects \cite{ARAUJO201298}. Hence, a correct emergent description of the three cases should be $(\eta, \psi')$, $(\eta, \psi'')$ and $(\eta, \psi''')$. The geometry part is the same, $\eta$, while the matter part, $\psi, \psi',\psi'',\psi'''$, being sensitive to number, location and orientation of the defects, is each time different\footnote{Clearly, these defects carry with them some strain, that makes the membrane a little deformed as compared to the pristine case, and introduce the emergent $A_\mu$ we already discussed. As said, this is not included in the present picture, nonetheless it is easy to do so by enlarging the ``matter content'' of the theory by including $A_\mu, A_\mu',A_\mu'',A_\mu'''$.}.

\begin{minipage}[t]{0.3\textwidth}
\includegraphics[width=\textwidth]{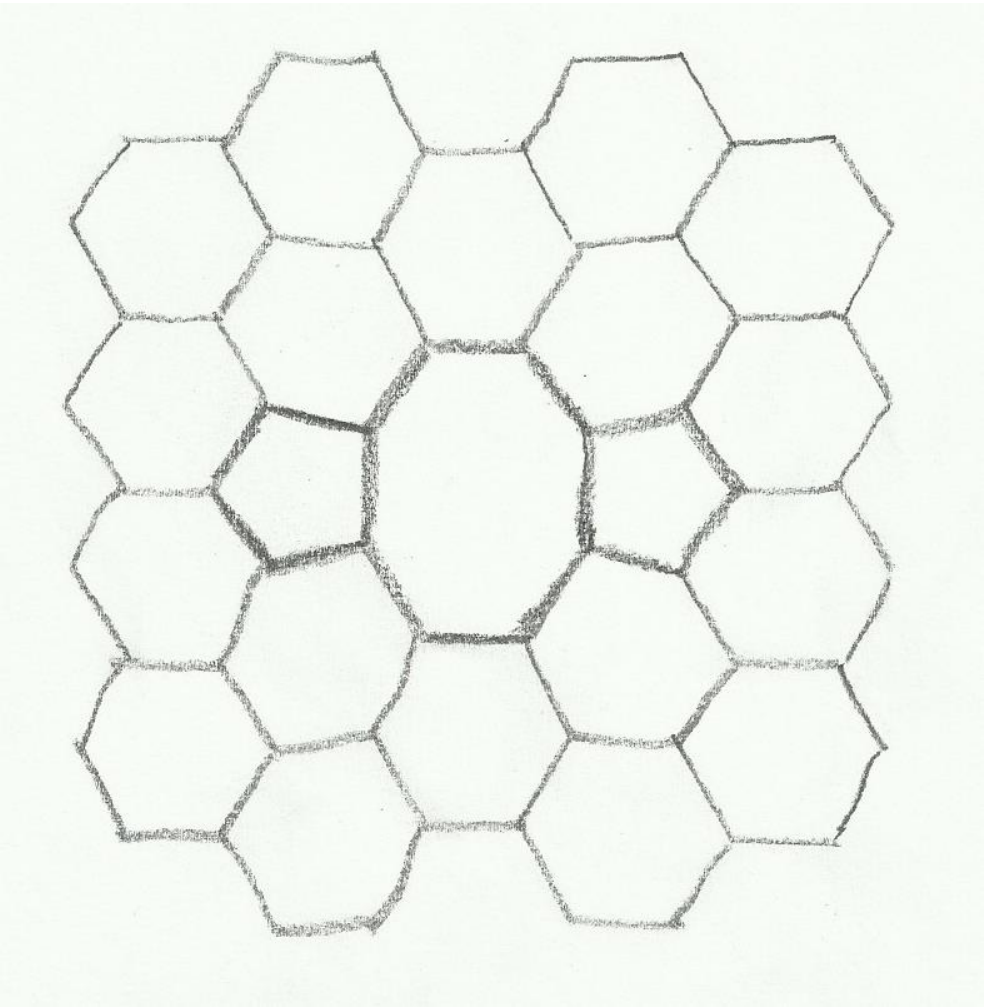}
\captionof{figure}{$(\eta, \psi')$}
        \label{585}
\end{minipage}
\begin{minipage}[t]{0.3\textwidth}
\includegraphics[width=\textwidth]{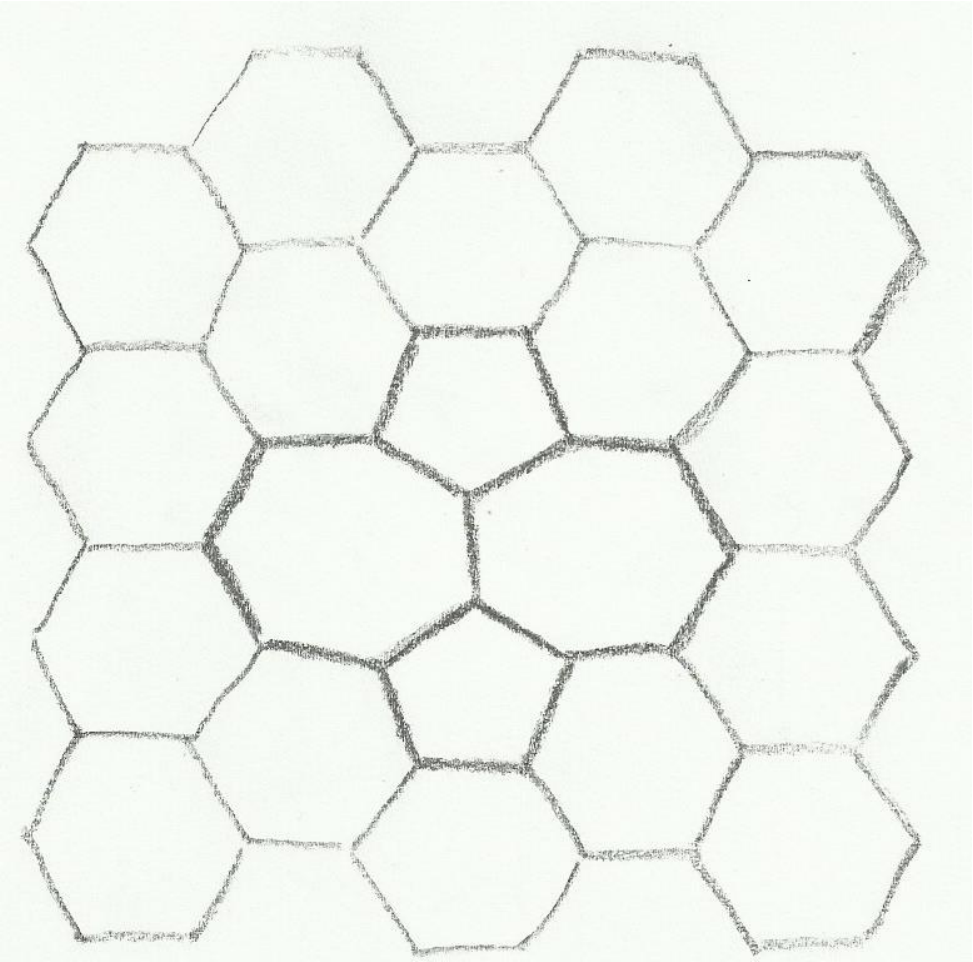}
\captionof{figure}{$(\eta, \psi'')$ STW }
        \label{STW}
\end{minipage}
\begin{minipage}[t]{0.3\textwidth}
\includegraphics[width=\textwidth]{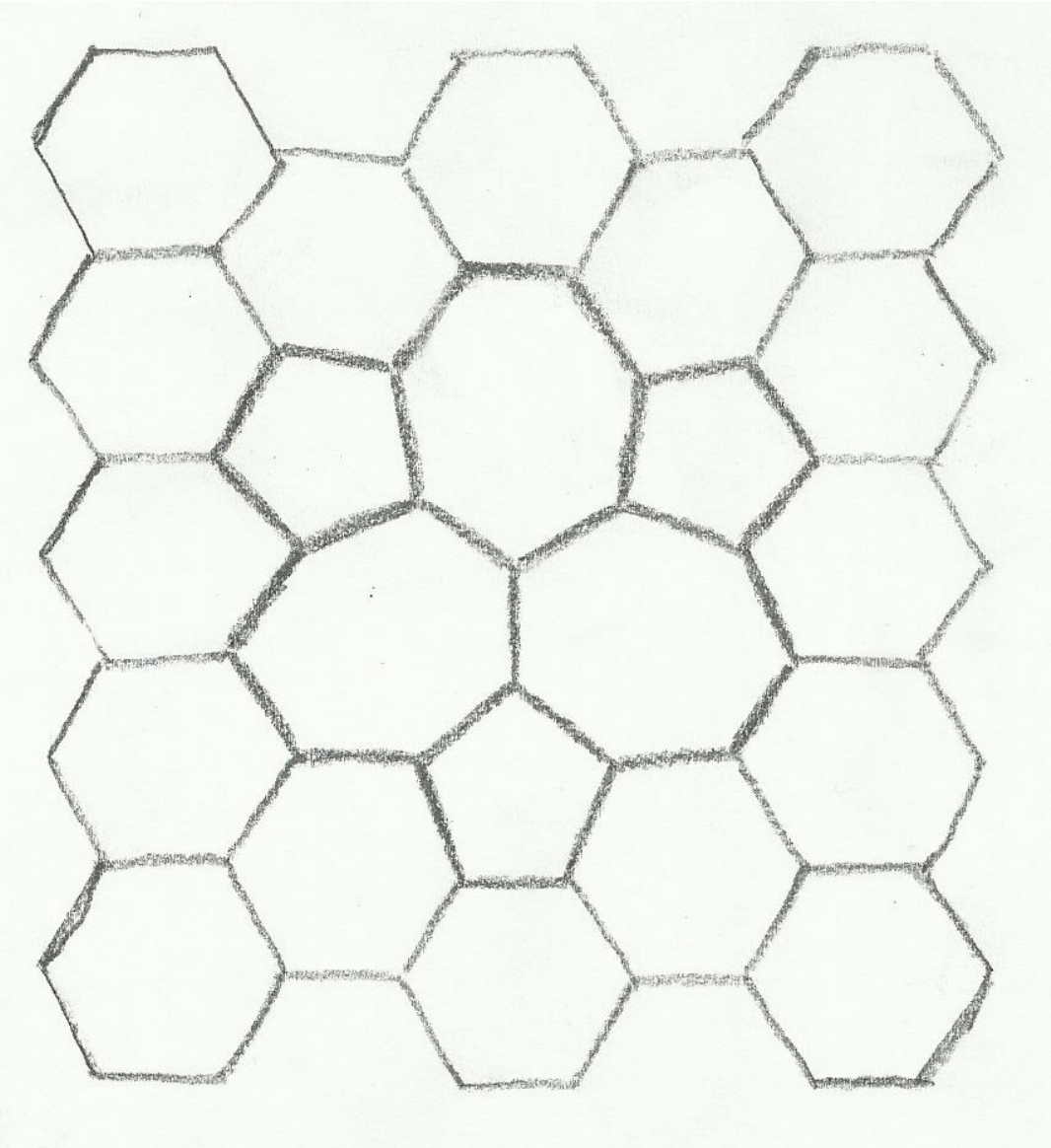}
\captionof{figure}{$(\eta, \psi''')$}
        \label{575757}
\end{minipage}

\begin{eqnarray}
                            &               Xons          \; \;       &  (\eta, \psi')  \;  \;  {\rm or} \nonumber \\
  (\eta, \psi)  \rightarrow & (g^{(a)}_{\BH}, \psi_{\BH}) \rightarrow &  (\eta, \psi'') \;  {\rm or} \label{evapscheme}\\
                            &                                         &  (\eta, \psi''') \nonumber
\end{eqnarray}

This scenario should then reproduce, with Dirac materials, the reshuffling of the fundamental degrees of freedom of the $X$ons (carbon's electrons, ions and photons, in the example of graphene), supposed to be at work during the evaporation as described in the previous Section. A schematic view is reported in (\ref{evapscheme}).

\subsection{The importance of analogs}

The field of analogs is, by now, an old and well established one \cite{Volovik:2003fe, Barceló2011}. Furthermore, due to our deeper understanding and experimental control of condensed matter systems, it is becoming increasingly popular to reproduce aspects of fundamental physics in analog systems. Examples are the cited Hawking phenomenon on graphene \cite{hawkinggrapheneplb,hawkinggrapheneprd} and in Bose-Einstein condensates \cite{Steinhauer:2015saa}, gravitational and axial anomalies in Weyl semimetals \cite{Gooth:2017mbd}, and more (for a recent result see, e.g., \cite{Ulf_LeonhardtPRL2019}). Actually, that does not apply only to condensed matter, as can be seen in the interpretation of hadronization in high energy collisions as a Unruh phenomenon \cite{Castorina:2007eb} (see also \cite{Castorina:2008gf} and \cite{Castorina:2014fna}).

Despite that, the importance of analogs for fundamental theoretical physics is still not appreciated. This is witnessed by the ongoing philosophical debates (see, e.g., \cite{Dardashti2016}), and by the fact that theorists look down at experiments on analogs as mere \textit{divertissements}, and never use any of those results to fix any fundamental theory. The rest of this paper is devoted to discuss why this is wrong, and to advocate the necessity to raise the status of analogs to the indirect experimental tests of fundamental theories, otherwise not-testable.

First, let us say that \textit{analogy}  (from the Greek $\alpha \nu \alpha$ -- $\lambda o \gamma o \varsigma$, , equal -- relation), fully qualifies to be a member of an important family of \textit{relations} among different descriptions, or different theories, or else, whose scope is that of a unified view, and points to an underlying invariant structure of nature.

The members of the family we could identify are four
\begin{itemize}
  \item \textit{Symmetry}. Here we have a theory, with action ${\cal A} (\Phi)$, mapped into itself
  \begin{equation}\label{symmetry}
    {\cal A} (\Phi) \to {\cal A} (\Phi') = {\cal A} (\Phi)
  \end{equation}
when $\Phi \to \Phi'$ is a mapping with some algebraic (Lie group, usually, but not escluseively) structure. The advantages are two. One is that there are conserved charges, pointing, in the case of spatiotemporal symmetries, to certain properties of spacetime (e.g., homogeneity and isotropy). The second advantage is that, as long as $\Phi$, $\Phi'$, ..., $\Phi''$ are all given by the same mapping, we can use any of the different descriptions, ${\cal A} (\Phi)$, ${\cal A} (\Phi')$, ..., ${\cal A} (\Phi'')$, at our convenience, to describe the \textit{same} physics. The descriptions, albeit equivalent, can be dramatically different.
  \item \textit{Duality}. This time, in general a theory is mapped to a different theory, or to a different regime of the same theory
  \begin{equation}\label{symmetry}
    {\cal A} (\Phi) \to {\cal A}_D (\Phi_D) \neq {\cal A} (\Phi)
  \end{equation}
when $\Phi \to \Phi_D$. There might be some algebraic structure behind the mapping, think of the $SL(2,R)$ structure behind the electromagnetic duality of certain supersymmetric theories \cite{IORIO2000171,IORIO2001156}. There might be no structure at all. Surely, here one cannot identify conserved charges. This is an important departure from the case of symmetry, because it is not easy to define what is the fixed entity we are describing in different and related, but not equivalent fashions. For symmetry, one can say that the fixed (invariant) entities are the conserved (invariant) charges, and the various descriptions only differ because different variables are used. Here the entities even change their nature, namely, collective solitonic excitations are mapped into elementary excitations, and \textit{viceversa}. What this relation shares with symmetry is the possibility to use different descriptions, ${\cal A} (\Phi)$ and ${\cal A}_D (\Phi_D)$, at one's convenience. The descriptions are not equivalent, but we have a mapping that allows to translate results in one setting, into results in the other setting. This last occurrence, and the fact that elementary and collective excitations are interchangeable concepts, make one think that there could be an underlying dynamics, common to both (or to more than two) dual formulations, of fundamental entities, in terms of which everything at the higher level of complexity, is just collective/quasiparticle.
  \item \textit{Correspondence}. Although someone might classify this relation as some form of the previous one(e.g., sometimes Juan Maldacena calls his AdS/CFT correspondence \cite{Maldacena1999} a ``holographic duality'' \cite{Maldacena2005}), we better dedicate to it a separate entry. In fact, this relation is more loose than the previous two, as it might include both, in certain limits, but surely is not a map from a theory into itself (symmetry) (or into a dual description) that works all the time. All that is necessary to have a correspondence is that certain key quantities, such as quantum correlators, are related. Therefore, this relation is one step ahead in the generalization of the first two relations of this family. It shares with them that there is a \textit{criterion} behind the relation. The criterion can be inspired by physics (e.g., the conservation of energy, the electromagnetic duality, holography), and again it points to an underlying structure common to all descriptions.
  \item \textit{Analogy}. Here we move a bigger step further, but, according to Feynman (see later), we are still in the same family. We are in the presence of an analogy between different physical phenomena when the equations governing such phenomena are identical. Therefore, the formal developments of the theory of one of them are valid for the others, provided we pay due attention to the fact that: a) the symbols used in the equations have, in general, different meaning, and b) the boundary conditions for one system may make no sense for an analog system, or could be very hard to realize in practice. Here there is no theory turning into itself (symmetry). Perhaps one can look at that as a theory mapped into another theory, so some sort of duality or correspondence, but in any case there is no \textit{criterion} for this relation. That is, what is missing is a reason, based on some physical argument (like conservation of energy, of charge, etc.), or inspired by some speculation (like electro-magnetic duality, or holography, etc), for establishing this relation. Missing a criterion, we do not have a specific \textit{mapping}. We have, instead, a \textit{vocabulary} of the meaning of symbols, on both sides of the relation: fields, couplings, sources, etc.. It is then very difficult to identify any common underlying structure.
\end{itemize}

The apparent lack of a criterion to establish an analogy, and the absence of arguments in favor of a common underlying structure, is what probably made the misfortune of analogs among pure theorists. Theorists use with no limitations all sort of combinations, spontaneous breaking, deformations, etc., of the first three members of this family, but look down to the last member as a curious coincidence, but nothing more. They, usually, do not know what to do with this, and, do not include analogs as possible experimental ways to tests their models. On this we shall comment in the following, with the help of Feynman, who had a different attitude towards analogs.

\subsection{Feynman's lecture and lesson on analogs}

In this last part of the paper, we offer some reflections on a lecture given by Feynman titled ``Electrostatic Analogs'', available in \cite{feynman2006feynman}. This lecture is very famous, but we are not sure that it is equally well known...

By this we mean that, probably, everyone has heard of the title of the first Section there, saying ``\textit{The same equations have the same solutions}'', but Feynman, in that Section and in the whole lecture, offers much more than that sentence, and we doubt that this made any serious impact.

This is a pity, because Feynman's extraordinary intuition and freedom of thoughts has proved pivotal in disparate areas of physics. Quantum computing is a prominent example of his visionary power. Thus, let us summarize here what is in that lecture, and let us comment on it.

The first Section, whose title is so famous, deals with the possibility for a physicist to be able to retain a broad knowledge as opposed to the apparent necessity for specialization. The reasons in favor of the broad knowledge, mentioned by Feynman, all refer to some underlying unity: general principles that always apply (like conservation of energy); complicated phenomena based on an underlying quantum electrodynamics (taken here as a prototype of a fundamental theory); and finally, the ``[...] \textit{most remarkable coincidence: The equations for many different physical situations have exactly the same appearance}'' says Feynman.

He goes on with ``\textit{Of course, the symbols may be different -— one letter is substituted for another -— but the mathematical form of the equations is the same. This means that having studied one subject, we immediately have a great deal of direct and precise knowledge about the solutions of the equations of another}''.

We disagree only with the use of the word ``direct'' there, as ``indirect'' is probably more correct. That said, though, Feynman put it so clearly that it is really surprising that analogs are still looked down by (too) many. But let us keep going.

After having said what quoted above, in that Section he recalls the basic equations of electrostatics, summarized by
\begin{equation}\label{Poisson}
  \vec{ \nabla} \cdot (\kappa \vec{ \nabla} \phi ) = - \frac{\rho_{free}}{\epsilon_0} \,,
\end{equation}
with standard notation. The point he wants to make here is that ``[...] \textit{there are many physics problems whose mathematical equations have the same form. There is a potential ($\phi$)
whose gradient multiplied by a scalar function ($\kappa$) has a divergence equal to another scalar function ($\rho_{free}/\epsilon_0$)). Whatever we know about electrostatics can
immediately be carried over into that other subject, and vice versa}''.

There are then five beautiful Sections dedicated to five such different physical problems, governed by (\ref{Poisson}): The flow of heat (a point source near an infinite plane boundary) -- The stretched membrane -- The diffusion of neutrons (a uniform spherical source in a homogeneous medium) -- Irrotational fluid flow (the flow past a sphere) -- Illumination (the uniform lighting of a plane).

For instance, the Section on the stretched membrane considers a thin rubber sheet stretched over a large horizontal frame.  Feynman shows that the displacement $u(x,y)$, obtained by pushing the membrane up in one place and down in another, obeys
\begin{equation}\label{PoissonMembrane}
  \nabla^2 u (x,y) = - \frac{f (x,y)}{\tau} \,,
\end{equation}
where $f$ is the force with which we push and pull, and $\tau$ is the surface tension. This is just the equation (\ref{Poisson}), with $u$ corresponding to $\phi$, and $f/\tau$ corresponding to
$\rho_{free} / \epsilon_0$ (here $\tau$ is constant, that corresponds to constant $\kappa$ there).  Therefore, having some fixed values of $u$ in some places is ``[...] \textit{analog of having a definite potential at the corresponding places in an electrical situation. So, for instance, we may make a positive `potential' by pushing up on the membrane with an object having the cross-sectional shape of the corresponding cylindrical conductor}''.

Here Feynman says that not only one can use electrostatics (a ``fundamental'' theory) to solve the elasticity (an ``emergent'' theory) problem, but also the reverse is true! He says: ``\textit{Various rods and bars are pushed against the sheet to heights that correspond to the potentials of a set of electrodes. Measurements of the height then give the electrical potential for the electrical situation. The analogy has been carried even further. If little balls are placed on the membrane, their motion corresponds approximately to the motion of electrons in the corresponding electric field. One can actually watch the `electrons' move on their trajectories. This method was used to design the complicated geometry of many photomultiplier tubes (such as the ones used for scintillation counters, and the one used for controlling the headlight beams on Cadillacs)}''.

What strikes here is that one can design specific shapes of the membrane, corresponding to electric potentials very difficult to obtain, and then let the physics of the membrane follow its paths, governed by equation (\ref{PoissonMembrane}). The results are valid for the other system, governed by the same equation, just with different meaning of the symbols, (\ref{Poisson}). Any further warning makes no sense. Actually, as noticed, Feynman says that one gets ``direct'' knowledge of system $A$, by studying the analog system $B$. We would say ``indirect'', instead, but that is the only place where we differ (and it is only a matter of wording). The fact is that there is no reason whatsoever for doubting that the results obtained with $B$ would be valid for $A$, and \textit{viceversa}.

Our question here is: so, why we are not fully exploiting the analogs to acquire with accessible systems knowledge of unaccessible systems? Although it might seem that analogs are very popular these days, this popularity is still not what we advocate here. Analogs are popular as a sort of amusement, but they are not considered as experimental tests of the fundamental theories.

For instance, with reference to Section 2 of this paper, it would be truly important to settle the information loss vs conservation puzzle through experiments. Although some analogs can be proposed (and we offered one in Section 2), the general attitude is that, these not being true $\BH$s, they cannot be used as test-beds of the theory. We think the opposite, as said already in other occasions \cite{FrontiersIorio,IorioDICEhelios}.

On the other hand, Feynman question here is: ``\textit{Why are the equations from different phenomena so similar}''? This means to look for the criterion behind the analogy, we absence of which we have pointed to in the discussion about the family of four relations. By evoking the  ``\textit{underlying unity of nature}'', Feynman starts elaborating on the possibility that  ``\textit{everything is made out of the same stuff, and therefore obeys the same equations}''. Then concludes that ``\textit{[...] the thing which is common to all the phenomena is the space, the framework into which the physics is put. As long as things are reasonably smooth in space, then the important things that will be involved will be the rates of change of quantities with position in space. That is why we always get an equation with a gradient. The derivatives must appear in the form of a gradient or a divergence; because the laws of physics are independent of direction, they must be expressible in vector form. The equations of electrostatics are the simplest vector equations that one can get which involve only the spatial derivatives of quantities. Any other simple problem —- or simplification of a complicated problem —- must look like electrostatics. What is common to all our problems is that they involve space and that we have imitated what is actually a complicated phenomenon by a simple differential equation.}''

Eventually, simply on the basis of what just quoted, he arrives at other questions, leading straight to the $X$ons that make (nearly) everything: ``\textit{Are they} [the electrostatic equations, ed] \textit{also correct only as a smoothed-out imitation of a really much more complicated microscopic world? Could it be that the real world consists of little $X$ons which can be seen only at very tiny distances? And that in our measurements we are always observing on such a large scale that we can’t see these little $X$ons, and that is why we get the differential equations}?''

Amazingly, the fact that systems $A$, $B$, $C$, etc are all analog, in a smoothed, approximated (emergent, in today's jargon) fashion of a system $D$ (electrostatics), that surely was deemed elementary at that time, lead Feynman look the way a physicist looks, i.e. in a democratic fashion: all systems, including $D$, are emergent descriptions of one more fundamental thing.

Although Feynman took brave steps into including space into the picture, as a fundamental criterion for the analogy, he really meant the properties of space (isotropy, in this case). So he only made half of the journey of Section 2 of this paper. That is, Feynman space is not really made of $X$ons: they need be there, but they do not make space itself.

Thus the last part of his comments, where he throws negative statements on the finding of a consistent theory of $X$ons, should be now confronted with explanations of the fundamental world where space also dramatically changes its nature at the ultimate $X$ level. For instance, Conne's noncommutative geometry \cite{DBLP:books/daglib/0076876}, appears to reproduces the \textit{standard} Standard Model of particle physics.

\section{Conclusions}

We elaborated on the hypothesis that both matter, and the geometry where it lives, emerge from the dynamics of more fundamental objects, as can be inferred from the Bekenstein bound on the entropy of any system.

A natural consequence is that, at our energies, fields and spaces are generally entangled entities. This leads in \cite{ACQUAVIVA2017317} to the noticeable result that $\BH$ evaporation is unavoidably a nonunitary process, hence with information loss. The final stage of the famous Page curve is then analyzed in this framework, and the conclusion is that there is always a nonzero entanglement entropy associated to the final products of the evaporation.

As this description should apply also to standard inequivalent representations of QFT, we evoke here a fascinating possibility for the tilde degrees of freedom of TFD. The question is whether such mirror images of the physical degrees of freedom could be interpreted as the way emergent fields see the degrees of freedom of geometry with which they are entangled.

We then suggest how certain scenarios of the quasiparticle picture could find analog realizations in condensed matter. In particular we propose graphene and graphene-like materials as possible analogs, with the necessary reshuffling of the fundamental degrees of freedom, between geometry and fields, at work during $\BH$ evaporation. In the example of graphene, the fundamental constituents, that we call $X$ons, are the carbon's electrons, ions and photons. A schematic view of how the evaporation mechanism should be reproduced with these systems is discussed.

As analogs popped up, we took the chance to discuss them in general. First by trying to include analogy into a family of relations (symmetry, duality, correspondence, analogy), and then by scrutinizing certain intriguing arguments of Feynman that elevate analogs from mere curiosities, to reliable tests of fundamental theories. Feynman's arguments point to the existence of fundamental constituents, that we see as predecessors of the $X$ons stemming from the Bekenstein bound, although the latter make also space not only matter.

The discussion of this paper adds even more reasons to construct an experimental facility, where to systematically test fundamental theories with analogs, as already advocated in various occasions \cite{FrontiersIorio,IorioDICEhelios}.

\section*{Acknowledgments}

The author is indebted to Giovanni Acquaviva and Martin Scholtz for their collaboration on these matters (and geometries), and for managing to bring the $X$ons (aka ``dots'') under the judgment of computations. He also thanks Georgios Lukas-Gerakopulos and Pablo Pais for many interesting and fruitful discussions.

\section*{References}

\bibliography{proceedings_biblio2}{}
\bibliographystyle{iopart-num}

\end{document}